\documentclass{mem}
\usepackage{natbib}\usepackage{txfonts}\usepackage{balance}
\usepackage{graphicx}

\usepackage[a4paper,breaklinks]{hyperref}
\idline{84}{1066}
\begin{document}
\def\teff{$T\rm_{eff }$}
\def\kms{$\mathrm {km s}^{-1}$}

\title{Calibration of stellar and atmospheric models using the Hyades
}

   \subtitle{}

\author{Taisiya G. \,Kopytova\inst{1,2}, Wolfgang \,Brandner\inst{1}, Siegfried \,R\"{o}ser\inst{3}, Elena \,Schilbach\inst{3} \and Nicola \,Da Rio\inst{4}}

  \offprints{T. Kopytova}

\institute{
Max-Planck Institut f\"{u}r Astronomie, K\"{o}nigstuhl 17, 69117 Heidelberg, Germany,
\email{kopytova@mpia.de}
\and
International Max-Planck Research School for Astronomy and Cosmic Physics at the University of Heidelberg, IMPRS-HD, Germany
\and
Astronomisches Rechen-Institut, Zentrum f\"{u}r Astronomie der Universit\"{a}t Heidelberg,
M\"{o}nchhofstr. 12-14, 69120 Heidelberg, Germany
\and
European Space Research and Technology Centre, Keplerlaan 1, 2200 AG
Noordwijk, Netherlands
}
\authorrunning{Kopytova et al.}

\titlerunning{Calibration of stellar and atmospheric models using the Hyades}

\abstract{Calibration and benchmarking of evolutionary and atmospheric models is essential for the study of
low-mass stars and brown dwarfs, as for isolated objects these models are the only way to determine
basic parameters like mass and age. The Hyades star cluster with an age of around 625 Myr and located
at the distance of $\sim$ 45 pc, is the most accessible cluster in the solar neighborhood.
R\"{o}ser et al. (2011) establish a list of 724 likely Hyades members.
Using available literature data (HST, HIPPARCOS, WDS, Patience et
al. 1998, Mermilliod et al. 2009, Morzinski 2011) and our own lucky imaging observations with AstraLux at the 2.2m telescope in Calar
Alto, we establish a single-star sequence containing 255 Hyades members spanning the mass range $\sim$ 0.2--1.5 solar masses . 
This sequence is used for testing and calibration of various existing stellar and atmospheric models (PADOVA, DARTMOUTH, BCAH, BT-Settl).
}
\maketitle{}

\section{Introduction}

The determination and knowledge of basic physical parameters of stellar and sub-stellar sources provides the basis for astrophysics.
Observations allow us to determine distance \citep[using parallaxes;][]{dl12,fah12}, age \citep[through membership in clusters or moving groups;][]{per98},
mass \citep[orbital parameters in binary systems;][]{kon10}, and radius \citep[transits;][]{joh10} for these objects.
But when dealing with isolated stellar objects, we have to rely on models to determining mass and age.
While stellar and atmospheric theoretical models are rapidly evolving, we need a powerful tool to calibrate them.
Open clusters can be a good candidate for this role, since they contain many coeval objects of the same chemical composition, thus avoiding the problems of small number statistics.

The Hyades is the closest open cluster to the Sun (d $\sim$ 45 pc) and suffers little from interstellar extinction.
The estimated age of the Hyades is 625 Myr \citep{per98}.
\cite{roe11} report 724 likely members of the Hyades with a kinematic distance estimate for each member using the convergent point method.
Therefore, absolute magnitudes for each source can be calculated and placed on color-magnitude and color-color diagrams.
Since there are binary and multiple systems among the 724 likely Hyades members, the main sequence on the diagrams is scattered. 
That is why, in this work, we use AstraLux lucky imaging observations and results from the literature to introduce a single star sequence of the Hyades.
Together with over-plotted theoretical isochrones, it allows us to test stellar and atmospheric theoretical models.

\section{Observations}
\subsection{AstraLux observations}
We observed 196 Hyades members using the lucky imaging camera AstraLux mounted on the 2.2m telescope at the Calar Alto Observatory in Spain.
Observations were carried out during two runs in November 2011 and December 2012.
Giving the size of the mirror of the telescope, the angular resolution is $\sim$ 0$\farcs$1 at 900 nm,
resulting on a typical separation of 5 AU that can be resolved for a Hyades member. 
36 targets were identified as candidates to binary or multiple systems (Fig. \ref{fraction}).

\subsection{Literature data}
\begin{figure}
\resizebox{\hsize}{!}{\includegraphics{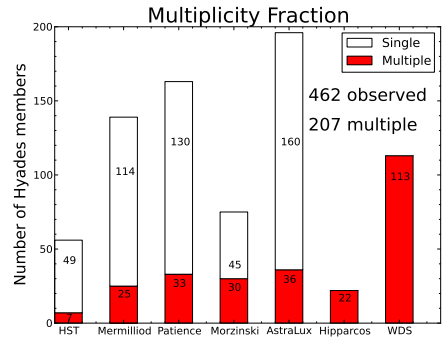}}
\caption{\footnotesize
The multiplicity fraction in the Hyades.
Combined data using Mermilliod et al. (2009), Patience et al. (1998), Morzinski (2011), AstraLux observations, and HST, HIPPARCOS and WDS catalogs.
The shaded part of each bar is a binary/multiple fraction from each survey. In total, 462 systems are observed and 207 candidates to binary/multiple systems are detected.
}
\label{fraction}
\end{figure}

We combined previously published Hyades data, which include spectroscopic observations \citep{mer09},
speckle binary search \citep{pat98}, adaptive optics observations at the Subaru telescope \citep{mor11}, and information
from HST, HIPPARCOS and WDS catalogs. The results are summarized in Fig. \ref{fraction}.
\vspace{12pt}

In total, 462 Hyades members were observed or found in the literature, 207 of these were identified as candidates to binary or multiple systems.
The targets that have at least one observation but does not reveal binarity/multiplicity (255 stars) are included to the single star sequence which is used in the further model testing.
It should be noted that some of the "single" stars still could be unresolved binaries.

\begin{figure}
\resizebox{\hsize}{!}{\includegraphics{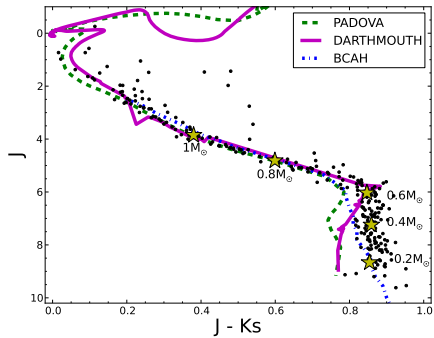}}
\caption{\footnotesize
The single star sequence of the Hyades (black dots) with over-plotted PADOVA (dashed line), DARTMOUTH (solid line) and BCAH (dash-dotted line) isochrones on J vs. J - Ks diagram.
}
\label{cmd_1}
\end{figure}

\begin{figure}
\resizebox{\hsize}{!}{\includegraphics{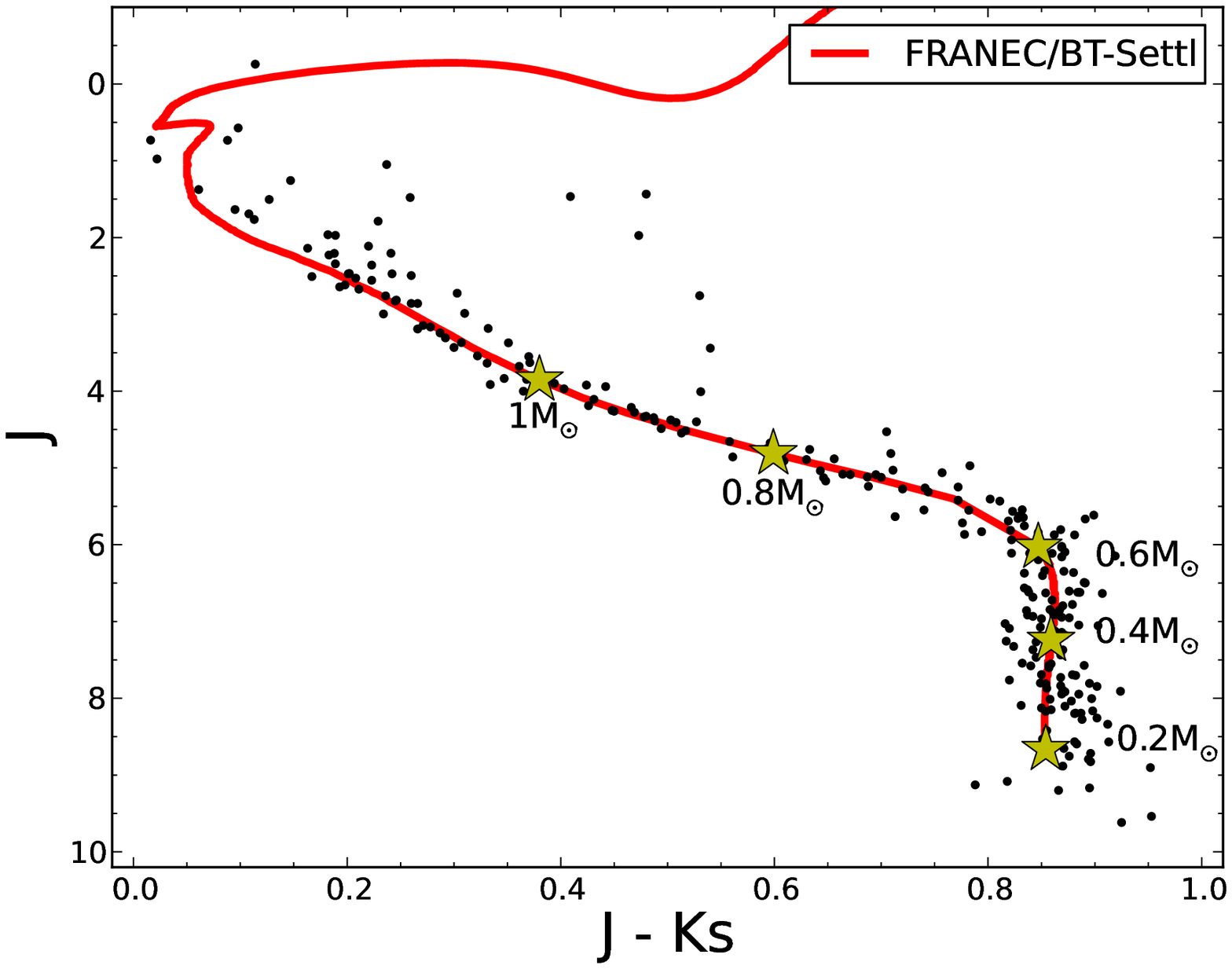}}
\caption{\footnotesize
The single star sequence of the Hyades (black dots) with the over-plotted FRANEC/BT-Settl isochrone (solid line) on J vs J - Ks diagram.
}
\label{cmd_2}
\end{figure}

\begin{figure}
\resizebox{\hsize}{!}{\includegraphics{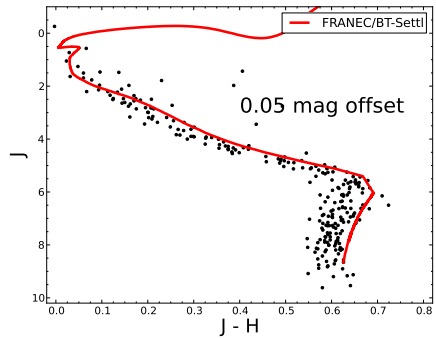}}
\caption{\footnotesize
The single star sequence of the Hyades (black dots) with the over-plotted FRANEC/BT-Settl isochrone (solid line) on J vs. J - H diagram.}
\label{cmd_3}
\end{figure}

\section{Model testing}
We place the identified single Hyades members on J vs. J - Ks color-magnitude diagram.
Additionally, we over-plot theoretical isochrones based on PADOVA \citep{mar08}, DARTMOUTH \citep{dot08} and BCAH \citep{bar98} models (Fig. \ref{cmd_1}).
We also use the Tool for Theoretical Data Analysis \citep[TA-DA;][]{rio12} to merge the stellar FRANEC model \citep{tog11} with the atmospheric BT-Settl model \citep{all11}.
As we can see, most of the models have troubles to describe the sequence at the lower-mass regime or around the "knee" ($\sim$ 0.6 M$_{\odot}$).
On the other hand, the combined FRANEC/BT-Settl isochrone describes the observed main sequence of the Hyades very well, even fitting the "knee" part (Fig. \ref{cmd_2}).

However, when trying to over-plot the FRANEC/BT-Settl isochrone with the Hyades single star sequence on J vs. J - H color-magnitude diagram, we notice an offset of about 0.05 mag (Fig. \ref{cmd_3}). This effect could be explained by the suggestion that opacities of alkali and hydride lines are still missing in the model (France Allard; private communication).

\section{Conclusions}
In this work, we reported a multiplicity census of the Hyades likely members resulting on identification of 207 possible binary and multiple systems out of 462 observed targets.
This allowed us to create the single star sequence containing 255 Hyades members and apply it to testing stellar and atmospheric models.
We found that the merged FRANEC/BT-Settl isochrone matches the sequence very well on J vs. J - Ks color-magnitude diagram but shows the offset of 0.05 mag on J vs. J - H.
The latter could be explained by missing theoretical opacities of alkali and hydride lines in the model.

We have shown that open clusters, in particular, the Hyades is a powerful tool to testing stellar and atmospheric models.

\begin{acknowledgements}
Authors are grateful to Katie Morzinski who allowed to use results of her PhD dissertation that contributed significantly to this work.
We thank France Allard for comments about atmospheric models.
We are grateful to Anatoly Piskunov for providing a compilation of theoretical isochrones.
And additionally, we would like to thank Betrand Goldmann, Elena Manjavacas, Joshua Schlieder and Niall Deacon for useful discussions.
\end{acknowledgements}

\vspace{12pt}

\bibliographystyle{aa}

\end{document}